# Power-law Distributions in Information Science – Making the Case for Logarithmic Binning

Staša Milojević

*School of Library and Information Science, Indiana University, Bloomington 47405-1901. E-mail: smilojev@indiana.edu*

**We suggest partial logarithmic binning as the method of choice for uncovering the nature of many distributions encountered in information science (IS). Logarithmic binning retrieves information and trends "not visible" in noisy power-law tails. We also argue that obtaining the exponent from logarithmically binned data using a simple least square method is in some cases warranted in addition to methods such as the maximum likelihood. We also show why often used cumulative distributions can make it difficult to distinguish noise from genuine features, and make it difficult to obtain an accurate power-law exponent of the underlying distribution. The treatment is non-technical, aimed at IS researchers with little or no background in mathematics.**

## Introduction

Information science (IS) is replete with distributions that can be characterized as a power law. Examples include author productivity (Egghe, 2005; Lotka, 1926; Pao, 1986), citations received by papers (Price, 1965; Redner, 1998), scattering of scientific literature (Bradford, 1934; Nicolaisen & Hjørland, 2007), and collaborative tagging behavior (Golder & Huberman, 2006). These constitute only a subset of empirically found distributions that follow a power law (an excellent overview of power laws and processes that can lead to these distributions is given by Newman (2005)).

Here we focus on some aspects of power law functions that are relevant to IS researchers. Mostly, we want to provide a method, *logarithmic binning*, that will help researchers recognize the presence or absence of power laws in their data. Descriptions will refrain from using mathematical formalism in order to make it useful for those who do not have mathematical or physical sciences training.

While detailed technical reviews of power laws exist in recent literature (e.g., Clauset, Shalizi, & Newman (2009) and Newman (2005)), these do not devote much attention to the *logarithmic binning* method. Binning is simply a procedure of *averaging* the data that fall in certain ranges of values (bins), and here we use it to "beat" the statistical noise and thus reveal the trends in the data. Binning is logarithmic, meaning that the bins have equal sizes in logarithms, which is, as we will see, a natural choice for a power law. Logarithmic binning is given some consideration (especially as a vast improvement over unbinned representations) in Adamic's (n.a.) online tutorial, but in it most attention is given to Pareto's cumulative distribution, which, we will argue, is not always a better alternative.

Power-law distributions can mathematically be represented with power-law functions:

$$y = cx^{-a},$$

where $a$ is the power law exponent, and $c$ an overall scale, or normalization. Power-law functions are monotonous, which means that when $x$ changes, $y$ either only decreases or only increases. When power laws are used to describe distributions, the exponent *a* is typically positive, meaning that when *x* increases, the *y* value decreases. Qualitatively this means that objects or events with a high value of some quantity are typically rare (there are few very prolific authors, very large cities, etc.). Power law distributions lead to phenomena such as the 80:20 rule. This rule, also known as the Pareto principle, was conceptualized by J. M. Juran, and states that 20% of causes lead to 80% of phenomena. It should be stated that this exact ratio (80:20) corresponds only to one specific value of the power-law exponent, $a = 2.16$ (calculated from Newman (2005), eq. 29).

Sometimes, certain distributions are described as scale-free in addition to being power law. This means that an increase by a certain factor at any value of *x* will produce the same decrease (or increase) in *y*. However, as Newman (2005) showed, the power law distribution is the only scale-free distribution, so the two expressions are synonymous, and we use only power law in this paper.

### Power-law distributions in IS

Historically, different IS phenomena have been described using the mathematical formalism of power law (Bookstein, 1976; de Bellis, 2009; Egghe, 1985). These power-law distributions have been given different names, although they are all related (Bookstein, 1990).

*Lotka's distribution (law).* Lotka's law is one of the best-known examples of a power-law distribution in IS, though not widely known outside the field. Its original formulation (Lotka, 1926) can be described in the following way: a large number of authors (*y* value) produces a small number of papers (*x* value), while very few produce many. Such a description, of course, is not precise since many distributions (not necessarily power law) exhibit such a property. More specifically, the power-law nature of Lotka's law can be illustrated in the following way. Let us take a power law exponent of 2, as suggested by Lotka, then, if there are 100 authors with one article, there should be 25 with two, 11 with three, and so on. Lotka's law is an example of a size-frequency distribution, which describes the number of sources with a certain number of items.

*Zipf's distribution (law).* Zipf's law comes from linguistics (Zipf, 1949). It is a rank-frequency distribution which describes the number of items in the source where sources are ranked in the decreasing order of frequency. Zipf's law originally applied to the number of times certain words have been used in a text, from the most frequent to the least frequent. This distribution is again a power law. We note that it is possible to construct the word occurrence distribution in the size-frequency manner as well. Thus, the Lotka "version" of the word frequency "law" would be that there are many words that appear rarely and few that are common, and that this distribution is a power law. So, words such as "the" or "a" would have large values of *x* in Lotka's distribution, while they would have low values of *x* in the case of Zipf's distribution. Moreover, it can be shown (Adamic, n.a.), that Pareto's *cumulative* distribution is equivalent to Zipf's rank-frequency distribution, with the *x* and *y* axes simply flipped. Quoting Adamic, one can see that "the phrase 'The *r*-th largest city has *n* inhabitants' is equivalent to saying '*r* cities have *n* or more inhabitants'". The first phrasing is Zipf's formulation, the second Pareto's.

*Bradford's distribution (law).* Bradford's law (Bradford, 1934) of literature scattering shows cumulative distribution of journals covering a specific subject. Mathematically, distributions of Bradford type are cumulative versions of Lotka's distribution (Burrell, 1991; Leimkuhler, 1967).

As has already been mathematically proved (Egghe, 1985, 2005), all of the above-mentioned distributions are equivalent. Basically these are all power law distributions that can come in "straight" form and are then known as Lotka's distribution, or in ranked or cumulative form and are then known as Zipf/Bradford/Pareto distribution. This work focuses on the Lotka form of power law distributions.

### Differences between formal power laws and empirical power laws encountered in IS

*Deviations from a power law.* First and foremost, empirical power laws (e.g., earthquake magnitudes, populations of cities, and number of telephone calls) are only approximations of mathematical power laws. Deviations of empirical distributions (including those in the IS) from a power law can be due to the fact that the underlying processes are more complex (and more difficult to formally describe) than a simple power law. In other words, one uses a power law as an approximation of true distribution. If certain data exhibit persistent systematic deviations from a power law it is worth trying to modify a simple power law to more accurately describe the data. Examples of such modifications are shown in Figure 1: (a) double power law (Csányi & Szendroi, 2004) in which there are two ranges with different power-law exponents (these appear like two straight lines on a log-log plot), (b) Pareto type 2 (or Lomax) distribution (Burrell, 2008; Glänzel, 2007) which tends towards a constant (straight horizontal line) for small values of $x$, but then turns smoothly into a power-law for large $x$, (c) power law with an exponential cutoff (Newman, 2005), where power law holds for small values of $x$, but then turns smoothly into a declining exponential function for large $x$. The exponential, large-$x$ tail drops faster than the power law. Another example of a deviation from a pure power law is (d) a log-normal/power-law distribution (Milojević, 2010), where the power law holds for large $x$, but a log-normal distribution describes small $x$. The log-normal part appears rounded on a log-log plot, and can have a maximum (peak) value, unlike the regular power law, or the examples (a), (b) and (c), all of which are monotonously declining.

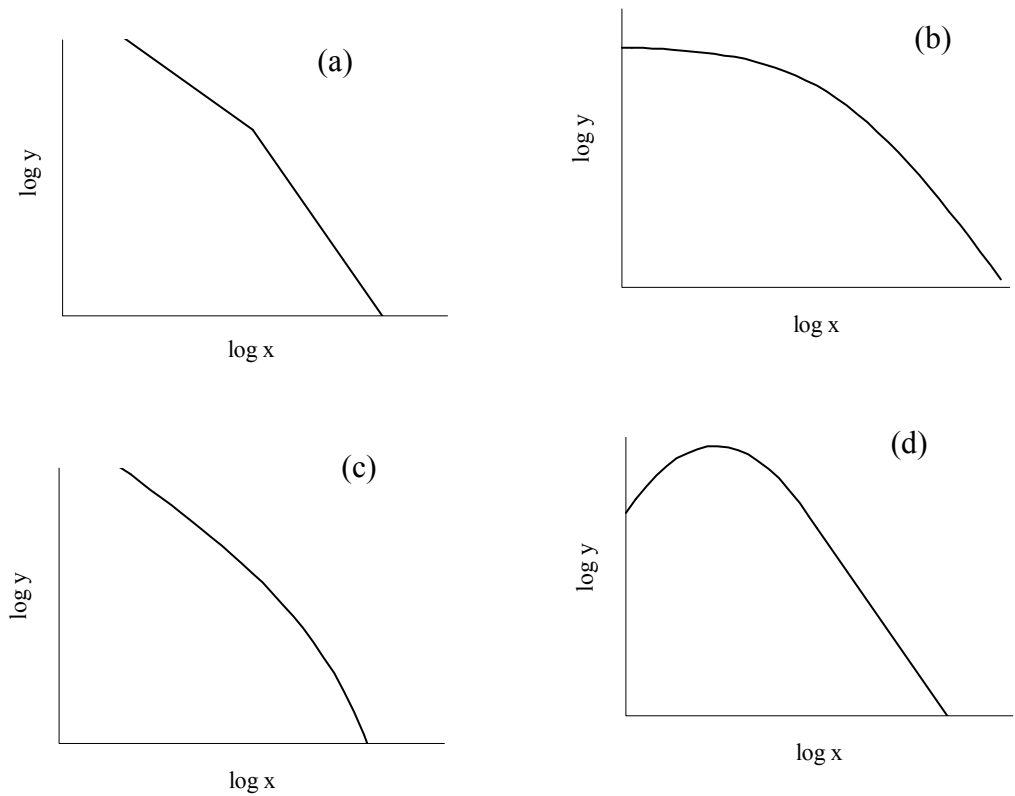

**Figure 1. Examples of deviations from a perfect power law. (a) Double power law is a power law with two different exponents. (b) Pareto II (or Lomax) distribution behaves as a constant function for small values of $x$, and as a power law for large $x$, with smooth transition. (c) Power law with an exponential cutoff behaves like a power law for small values of $x$, but drops more steeply (exponentially) for larger $x$. (d) Log-normal/power law composite has a log-normal distribution (which can have a peak) for smaller $x$, and a power law for larger $x$.**

Determining whether the data are actually drawn from a pure power law or not is beyond the scope of this work. Methods that can be used to establish this are given in Clauset et al. (2009). However, we note that it is fairly common to interpret the results in IS as power laws, even when it is obvious that they are not pure power laws. One reason for doing so is that power laws are easily described with basically one number, the exponent.

*Discrete vs. the continuous values of x.* Another important difference between a formal power law function (as, for example, described in introductory algebra textbooks) and its manifestation in IS is that the power law, as most mathematical functions, is defined (has some value *y*) for a continuous range of real numbers *x* (numbers that in addition to whole, "round" numbers include decimal numbers). This continuity is shown as a solid line in Figure 2a. This is not the case in real IS distributions, where *x* must be a whole, positive number (1, 2, 3, …).[1] In other words, power laws in IS are *discrete* (non-continuous); they are defined only when *x* is a natural number. To distinguish it from the continuous power law, we will refer to such *x* as *k*.

*Statistical nature of empirical power law distributions and the noisy tail.* Furthermore, in empirical power law distributions the *y* value is also a non-negative integer (0, 1, 2, 3, …) and therefore discrete (Figure 2c). We refer to such integer values of *y* as $y_k$. Taking again the example of Lotka's law with exponent 2 and normalization of 100, we formally expect 6.25 authors to have produced four papers ($100/4^2=6.25$). In reality, this has to be an integer, and we naturally expect it to be 6. However, empirical power laws are not just the "rounded" versions of theoretical power laws. This "rounding" doesn't capture the real, probabilistic nature of empirical distributions. In the above example, while we expect outcome to be 6, other values are also possible. The exact answer can be obtained by calculating a Poisson distribution with a mean of 6.25. From it one can see that, as expected, 6 is the most probable outcome (with a probability of 16%), but that 5 and 7 are similarly probable (15% and 14%, respectively). Other numbers (including zero) are possible too, but with decreasing probabilities. This probabilistic aspect leads to the *statistical noise*, which is always present, but is most obvious in the tails of discrete power law distributions, i.e., when the expected value of *y* is low. Again referring to the example of Lotka's law given previously, the expected number of authors that have produced 20 papers will be 0.25 (=$100/20^2$), but in reality this has to be a whole number. Calculating a Poisson distribution actually tells us to expect zero with a probability of 78%, one with 19%, two with 2%, etc. To summarize, in empirical power law distributions we typically find that $y_k$ is some large number for small $k$, and consequently the noise is imperceptible. However, at large $k$ the $y_k$ values fluctuate a lot and the noise becomes visible (Figure 2c). In tails $y_k$ are typically one or zero, with zeros becoming prevalent until we reach the last data point.

---

[1] Even in instances when the distribution may be modified in such way that *x* values do not represent integers, (for instance, if one constructs the productivity distribution, but assigns partial "credit' (<1) for authors who are not the sole authors of a paper), *x* still represents discrete events and is not continuous. Also, we note that *x* can in some distributions take a value of zero (for example, number of papers that had zero citations), but since this value cannot be represented in the power law formalism (except with some modifications), it is usually ignored.

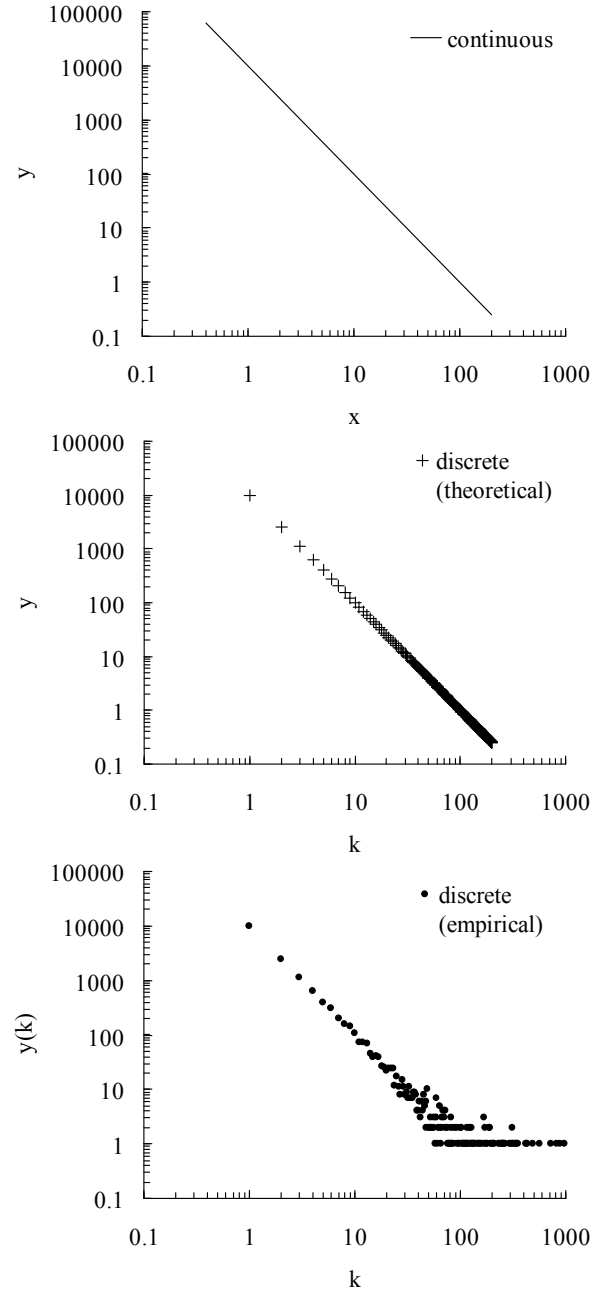

**Figure 2. Distinctions between different types of power law functions. All three functions have the same exponent ($a = 2$). (a) Continuous power law is defined for any value of *x* and appears as a perfectly straight line. (b) Discrete power law (i.e., Lotka's law) is defined only for natural numbers, which is why no points exist left of $k = 1$. Points get closer together for large *k* because the difference between the logarithms of consecutive numbers becomes smaller. Values of *y* are real numbers and points lie on a perfectly straight line. (c) Empirical discrete power law is a random "drawing" from the theoretical discrete power law. Panel shows one such drawing. Due to the statistical nature of the empirical distribution, deviations from (b) are present, especially for large *k* (the tail). Now $y_k$ must take integer values, which is why at large *k* we typically have either one or zero, leading to noticeably noisy appearance.**

Noise leads to deviations that are solely *statistical* or *probabilistic* in nature, and should not be confused with the (expected) deviations of empirical systems from theoretical models discussed previously. Distinguishing the two can be difficult, and the method we suggest here should aid in that respect.

Note also that the noise in the tail should not be confused with what is known as a "fat tail", despite the suggestive appearance. A fat tail is simply the property of *any* power law distribution with respect to a *normal* distribution. That is to say, in typical power laws the probabilities get smaller when $k$ increases, but not as drastically as for the normal distribution, where values far from the mean are extremely unlikely. So in comparison with the normal distribution, the power law tail has much higher values, i.e., appears "fat", especially in linear plots.

**Retrieving the information from noise using binning**

Typically in IS one presents distributions in plain, unbinned form (i.e., each data point is presented separately). As an example of this, we will produce a productivity plot using the bibliographical data of papers from the Nanobank database (Zucker & Darby, 2007). For each author we count the number of papers published. Next we count how many authors have a given number of papers. We show the number of papers and the number of authors on a log-log plot (Figure 3).

If the points appear to fall on a straight line, as expected for this kind of distribution, we interpret this as a power law. However, whether the power law actually holds in the tail is not obvious from such plain representation because of the statistical noise that dominates over a possible true trend[2]. To be able to verify the presence of the power law in the tail, or to see if any non-power law trends exist in it, this seemingly lost information should be retrieved in some way. More generally, what we aim to do is to use data in Figure 2c to reconstruct the underlying power law from Figure 2b. One such procedure, which we recommend, is the *partial logarithmic binning*.

Binning is a procedure of averaging the data that fall in specific bins, i.e., ranges of $k$. Averaging produces a more accurate answer as to what the real expected value of the function is. We take the bins to be of the same size when $k$ is given in logarithms, which is why this is *logarithmic* binning[3]. Finally, the binning that we perform is *partial*, because we bin the data only for $k$ larger than some number. That is to say, for small $k$ the data have very little statistical deviation and imposing the binning can unnecessarily smooth over some real trends. Furthermore, for small $k$ the logarithmic separation between consecutive $k$ can be larger than the binning interval itself, which can introduce unnecessary difficulties. For example, let us say that we have chosen a binning interval of 0.1 decades[4]. Thus for $k = 2$ and 3, the log-separation between them is log 3 – log 2 = 0.17, which is larger than the bin size, so we do not wish to start binning there. A procedure to select the place where to start binning is suggested later.

Another important application of binning is that it allows us to determine the underlying power law exponent of the distribution without bias. The power law exponent for the non-binned distribution is typically calculated in the following way: logarithms are calculated for both $k$ and $y_k$; a linear fit is obtained (usually using the least-square technique) such that the line passes closest to all points. In this technique, each data point has the same *weight* in determining of the fit. Since most data points have large $k$, and points with large $k$ are most affected by statistical fluctuation, the resulting fit will be biased by the noise. Fitting to logarithmically binned data naturally overcomes these biases and allows every range of the distribution to contribute equally to the fit. The result of this procedure is the improved ability to retrieve the underlying power law exponent.

---

[2] By "tail" we will consider the range of power law distribution where statistical fluctuation starts to become evident. The place where the tail begins will depend on the size of the system.

[3] Note, however, that to calculate the $y$ value corresponding to the bin ($y_B$), we take averages of plain $y_k$ values and not of logarithms of $y_k$.

[4] An interval where $x$ changes by a factor of 10 is called a decade.

### Binning procedure

Logarithmic binning is both practical and instructive, and becomes essential at large $k$ where the statistical noise becomes significant. On the other hand, data are best left unbinned for small $k$. If we are dealing with datasets containing in excess of ~10,000 data points, we can determine the starting point for binning ($k_b$) at the place where the statistical noise in $y_k$ becomes larger than some threshold $t$. If, for example, we choose $t = 0.02$ (2%), this will correspond to $k$ such that $y_k = (1/0.02)^2 = 2500$. In this example, bins 1 through $k_b$ will not be binned because they have $y_k$ values higher than 2500.

The size of the bin should be such that it does not "oversmooth" the data and any possible real trends, and yet not so small that the effects of smoothing are lost. We find 0.1 decades to be a good bin size for typical IS distributions (regardless of the total size of the dataset). This means that the interval between 10 and 100 will be split in 10 bins, as will the one from 100 to 1000, etc. A recommended range for bin sizes is between 0.05 and 0.2 decades.

For smaller datasets (with fewer than ~10,000 data points), the statistical error in $y_k$ will always be above the noise threshold of several percent. In such cases we may choose a starting point for binning ($k_b$) such that the separation between successive values (log $(k + 1)$ – log $k$) becomes smaller than the size of the bin. For example, if the bin size is 0.1, then the binning will apply for $k \geq 5$, since log 5 – log 4 = 0.097, which is smaller than 0.1.

Once the bin size and the starting point for binning are selected, for each bin we calculate the binned value ($y_B$) simply as the average of all $y_k$ values (1 through $n$) such that $k$ falls in the bin:

$$y_B = \frac{y_{k1} + y_{k2} + \cdots + y_{kn}}{n}.$$

Note that $n$ is the number of all integers that fall in the given bin, regardless of whether $y_k$ is zero or not. Note also that in the last bin $n$ needs to include all integers even if they are beyond our last data point (highest $k$ value).

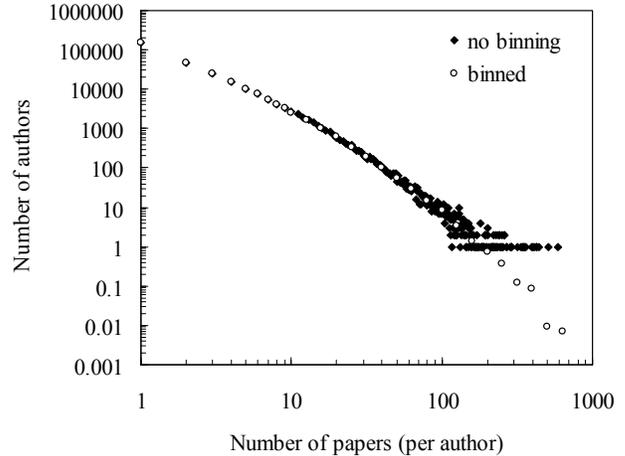

**Figure 3. Distribution of the number of papers per author from the NanoBank database. Points before binning are shown as black dots. Binned data are shown as open dots. Binning allows us to see that the Lotkaian power-law decline extends even for authors with extremely large number of papers.**

We illustrate the binning procedure using again the bibliographic data from NanoBank database of nanotechnology-related articles (Zucker & Darby, 2007). For the period 2000-04, the database contains 270,135 articles, authored by 294,456 authors. In Figure 3 the black dots show the plain, unbinned distribution of the number of authors who have published $k$ papers, i.e., the classical Lotkaian distribution. We see that the distribution is roughly a power law. Because of the very large total number of data points, the distribution stays smooth for relatively high $k$. However, at $k \sim 50$, the statistical deviations of black points start to become noticeable. At $k > 100$, we reach a point where we have very large deviations, and soon after the $y_k$ values become two, one or zero. Superimposed on the empirical data in Figure 3 are the logarithmically binned points (open circles). For $k < 10$ the data is not binned, so each category from 1 to 9 is kept with original values, and coincides with black points. For $k \geq 10$ the statistical fluctuations exceed 2% and the data are binned in bins of 0.1 decades. For example, $k = 15$, 16 and 17 have log $k = 1.18$, 1.20, 1.23, and they will all be placed in a bin centered at 1.2 (spanning the logarithmic range from 1.15 to 1.25). For $k = 14$, $\log k = 1.146$, so it will be placed in the preceding bin, while for $k = 18$, $\log k = 1.26$, and it will be placed in the following bin. One then averages the values of $y_k$ corresponding to $k = 15, 16$ and $17$, according to the above equation, even if some of them are zero (that is, $n$ is always 3 for this bin). Looking at the binned points in Figure 3, we see that while the unbinned data start to show increasing random fluctuations for $k > 50$, the binned points are still smooth. The most striking difference appears at the end of the tail. While the unbinned data end in noise, the binned points show that the smooth power-law decline of the number of authors with large number of papers *continues*. Of course, such binned *y* has to be interpreted as the expected value. For example, the binned data tell us that we should expect to have 0.1 authors who have published 300 papers, which can be interpreted that the expected number of authors who publish between say 295 and 304 papers is one.

The example shown in Figure 3 also helps us see any hidden trends. Binned points now indicate that the power-law trend is not perfect, but is slightly curved throughout. Indeed, a formal fit shows that the quadratic fit is a much better functional description of the trend than the linear fit. A quadratic function in log-log space corresponds to a log-normal function, one of the "deviations" of a power law mentioned previously, and discussed at length in (Milojević, 2010). A reasonably good fit is also obtained with a Pareto type 2 (Lomax) function described earlier.

How do we know that the trend uncovered from the noisy tail is indeed real? One way to establish this is using simulations. In Figure 4a we show a model double power law function that changes exponent at $x = 50$. At that point the slope changes from 2 to 1.1. Using this theoretical model we produce (simulate) data points drawn from this function, shown in Figure 4b. We see all the usual statistical fluctuations in the tail. Most importantly, we now see that it is hard to realize that there is an underlying double power law. That feature is lost in the noise and the distribution looks similar to one in Figure 2c, which is a single power law. Next, we apply logarithmic binning to these simulated data. The result is shown in Figure 4c. Binning starts at $k = 6$, and the bin width is 0.1 decades. Binned points remove most of the noise present in the data and reveal the existence of the double power law. This illustrates how binning reveals real features present in the data, some of which may be critical for understanding the problem being investigated.

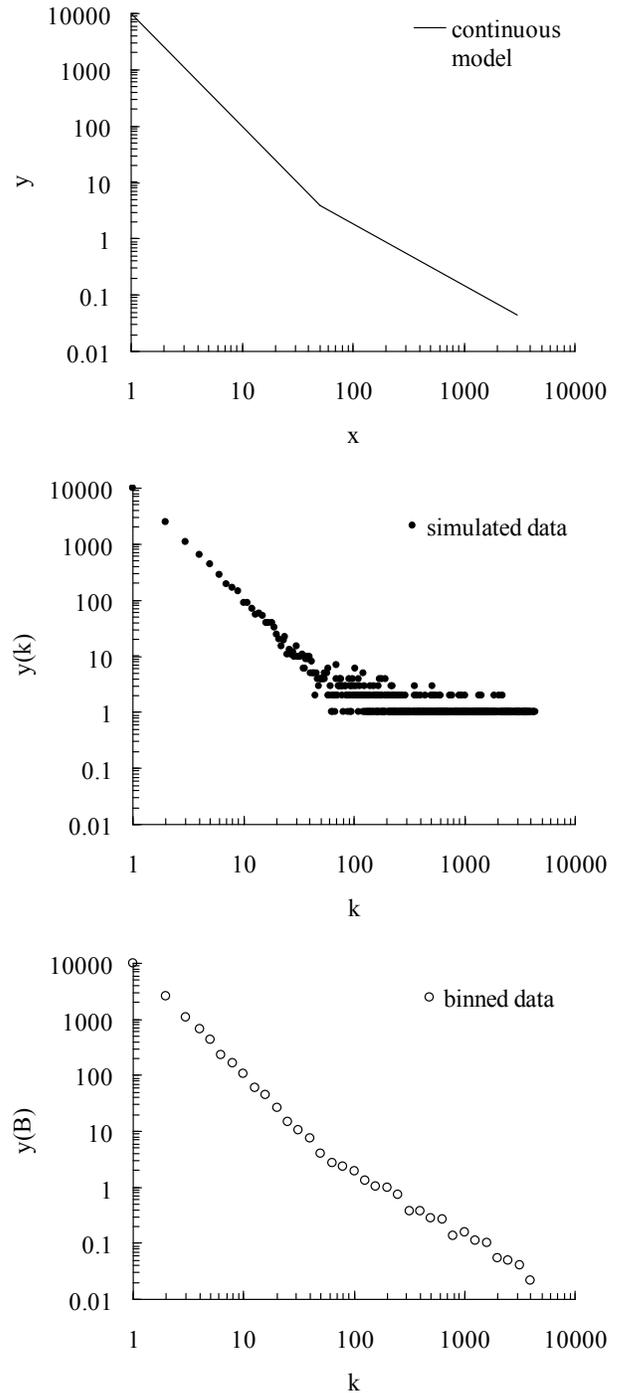

**Figure 4. Retrieval of real trends in the case of a double power law. (a) A composite continuous function, constructed from power laws with different exponents (changing from 2 to 1.1 at $x = k = 50$). (b) Simulated "data" drawn from a discrete version of the function shown in panel (a). The second power law is completely hidden in the noise. (c) Logarithmic binning of the "data" from panel (b). Binning retrieves the double power law structure that was hidden (but apparently not lost) in the noise.**

## Extracting the exponent

If a distribution resembles a power law, one typically tries to estimate its power-law exponent $a$. This enables a simple quantitative comparison with other studies. The power-law exponent can be easily visualized as a slope of the line in the log-log plot. Note that in cases where the power law inherently deviates from a perfect power law (as for example in Figure 3), the determination of the exponent $a$ will be sensitive to the range of $k$ used to obtain the slope. In such cases the slope can be quite different if fitted only to the core of the distribution (small $k$), or only to the tail.

Figure 2b shows a theoretical power law with a slope $a = 2$. However, if we were to use the ordinary least square method (without weighting) to obtain the slope from the "data" (Figure 2c), the result would be 1.31 (with all points included in the fit). This is a significantly biased result not reflected in the formal statistical error of the slope, which is only 0.15. The flattening of the slope is due to a large number of points in the noisy tail that lie above the underlying theoretical slope. One commonly used procedure to combat this bias is to exclude the data points from the tail, but this is a rather arbitrary and wasteful procedure. A much better alternative is to perform the ordinary least square fit on the logarithmically binned data. Applied to points from Figure 2c this method gives a result of $a = 2.10$, much closer to the expected value and entirely consistent with it (statistical error of the slope is 0.23).

One alternative to the least-square fitting is the maximum likelihood method. Newman (2005) provides a simple formula to obtain the power law exponent from the data (without any binning). However, that formula works only for continuous power law distributions (Clauset et al., 2009). Applying it to data that were drawn from discrete distributions produces altogether incorrect results. For example, applied to points in Figure 2c, it yields $a = 2.8$. For such distributions one must apply maximum likelihood methods for discrete distributions, which, unfortunately, cannot be expressed as a simple formula, but require numerical integration (Clauset et al., 2009). However, easy to use tables that show the relation between data sums and the exponent are given in Rousseau (1993)[5] and have been implemented in the software LOTKA[6] (Rousseau & Rousseau, 2000). Running LOTKA on data from Figure 2c gives the exponent of $a = 2.01$, very close to true value.

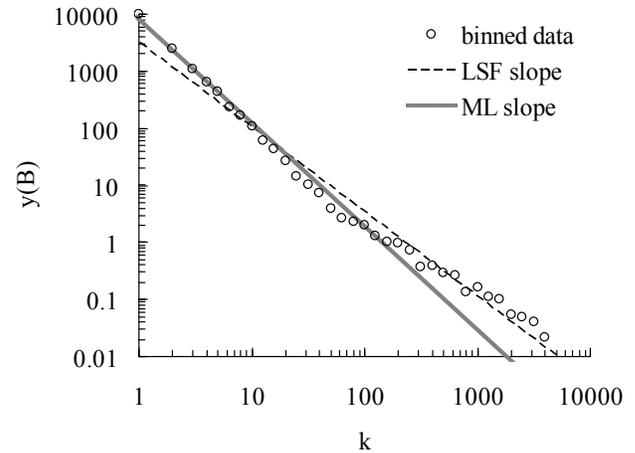

**Figure 5.** Extraction of power law exponents when the underlying function is not a true power law. Underlying distribution is actually a double power law from Figure 4. A least square fit to binned points produces a dashed line, while maximum likelihood method yields a shaded line.

While there may be little doubt that the maximum likelihood method is preferable to the least-square fit in cases where the underlying distribution is *known* to be a true power law, this is less obvious in real-world cases, where distributions deviate from power laws. For example, data in Figure 4b come from an underlying distribution that is a combination of two power laws ($a = 2$ and $a = 1.1$). Least square fit to logarithmically binned data produces an overall exponent of 1.49, shown with a dashed line in Figure 5. This exponent passes close to the binned data for the entire range of $k$. The exponent obtained using the maximum likelihood method and unbinned data is 1.84 and is shown with a grey line. This exponent describes data better than the least-square fit for smaller $k$ (before and around the place where exponent changes), but starts to deviate increasingly in the very tail. Which method is "better" may depend on the question we are asking. Here we show that the least-square fitting to logarithmically binned data may be instructive in some cases, and should be used in addition to the maximum likelihood methods, especially when we are not certain that the underlying distribution is a pure power law.

---

[5] Also reproduced in Egghe (2005).

[6] Available from http://www.cindoc.csic.es/cybermetrics/articles/v4i1p4.html.

Let us again emphasize that the least square fitting should never be applied to unbinned data. For example, if applied to the data in Figure 4b, it would give an exponent of 0.55, a result seriously affected by the noise in the tail.

## Logarithmic binning versus cumulative distribution

In this paper we present logarithmic binning as a way to reduce the effects of the statistical noise in empirical IS distributions. Cumulative distributions have been suggested as an alternative or even in "many ways a superior" method to logarithmic binning (Newman, 2005, p. 326). In cumulative distributions (again, the focus is on discrete empirical distributions), one shows the number of objects having a value larger than $k$, instead of the number of objects with value exactly $k$ as was the case so far. For example, for productivity distributions, $y_k$ when $k = 1$ would be the number of authors who have published one or more papers, and not those who have published exactly one paper. Value of $y_k$ when $k = 2$ are all authors with two or more papers, and so on. It is easy to see that each successive $y_k$ will be smaller than the previous one, i.e., cumulative distributions always fall. Consequently, the cumulative distribution is strictly monotonous, unlike the non-cumulative distribution where noise can make the function fluctuate up and down. This strict monotony leads to a smooth appearance of cumulative distributions with little or no *apparent* noise in the tail.

We need to emphasize that the many perceived advantages of cumulative distributions may not always be there. First, it is generally recognized that if a regular, non-cumulative function is a power law, than the cumulative function would also be a power law only with an exponent that is smaller by exactly one (Newman, 2005). What is often overlooked is that this applies only to continuous functions. A cumulative function of a discrete power law will result in a distribution that curves up for small values of $k$. For example, for a power law with exponent of 2, we expect the cumulative distribution to have the exponent of 1. However, in the discrete case the slope between points $k = 1$ and $k = 2$ will be 1.35. Exponent approaches the expected value of 1 only for larger $k$. This can again be illustrated using the data from Figure 2c, where we know the underlying exponent is 2. We show its cumulative version in Figure 6. We see that for small $k$ the cumulative points are steeper than the expected exponent of 1.

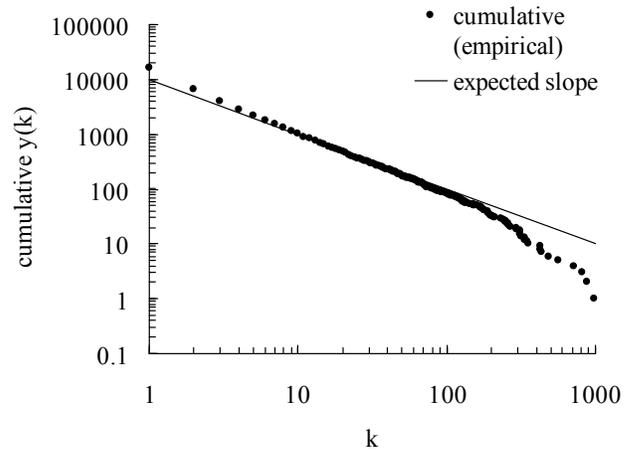

**Figure 6. Cumulative version of the power law "data" from Figure 2c. The function that generated the "data" has the exponent of exactly 2, which means that the exponent of 1 is expected for the cumulative distribution, shown as a line. Actual cumulative distribution deviates from the expected straight line at both ends, and is overall more steep.**

Our main objection to the use of cumulative distributions in IS is that they do not really eliminate the noise in the tail, they just make it more difficult to see it or to distinguish it from real features. Noise appears to be absent simply because the cumulative distribution is monotonous, and therefore smooth and "nicer looking". That the noise is still there we see again in Figure 6, by the increasing departure in the downward direction of cumulative points from the expected straight line. Departure from the expected slope starts around $k = 30$, but is difficult to see before $k = 100$. If we did not know what the true exponent of this distribution was, we would have a hard time realizing that the increased slope at large $k$ is not real but due to the noise. Not surprisingly, obtaining the least square fit to such cumulative distribution gives a steep slope with the exponent 1.23, a result that is further from the correct one than those produced by the logarithmically binned or maximum likelihood methods applied to non-cumulative versions discussed previously.

That cumulative distributions make noise and true features difficult to distinguish is even more striking when the underlying distribution is not precisely a power law, as is often the case in IS. In Figure 7 we show a cumulative version of the double power law distribution of Figure 4b. It appears very smooth. Dot-dashed lines show what one would naively expect the slopes to be, by taking the non-cumulative exponents and reducing them by one. The cumulative distributions look nothing like it. The double power-law character of the underlying function is very difficult to see. The main reason for this is that each point is "affected" by all of the points that come after it. So even in the range where the effects of noise are small ($k < 30$), and where even the straight data plot (without any binning) shows a clear power law behavior (Figure 4b), the cumulative representation leads to a curving. Also, we again see that noise is not really eliminated, but appears as a strong downward trend for large $k$.

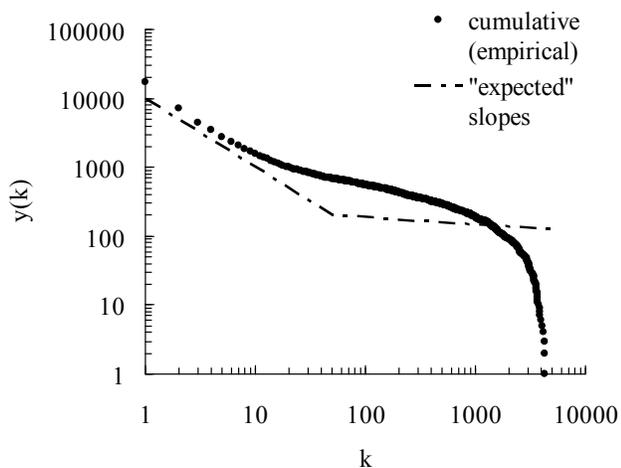

**Figure 7. Cumulative version of the double power law distribution from Figure 4.** One may naively expect that the power law character present in the original distribution will be preserved in the cumulative distribution, only with different exponents (dashed-dotted lines). Instead, even the left part of the distribution is "affected" by the right part, and the true underlying nature is difficult to see. The strong downward trend at large $k$, despite its smoothness, is not real but is the result of the noise.

## Conclusions

The statistical nature of sampling will always lead to the increasing noise in the tails of empirical distributions of the power-law type, regardless of the sample size. We show that applying the procedure of logarithmic binning allows researchers to better explore the functional forms of their distributions, especially in the tails. Binning also allows an unbiased power law exponent to be determined using the traditional least square method and without discarding the data in the tail. Getting the exponent from the binned data is recommended in addition to procedures such as the maximum likelihood method, which is preferable when the underlying distribution is known to follow a pure power law. For most IS applications we suggest binning that starts from $k = 5$, and using intervals of 0.1 decades.

While in many cases practical, we demonstrate that constructing cumulative distributions may not be the method of choice when one has (a) discrete distributions or (b) when the underlying distribution is not a perfect power law. Both of these characteristics are common in the distributions encountered in IS.

## Acknowledgments

I am grateful to Blaise Cronin and an anonymous referee for their comments and feedback. Certain data included herein are derived from NanoBank (Lynne G. Zucker and Michael R. Darby, NanoBank Data Description, release 1.0 (beta-test), Los Angeles, CA: UCLA Center for International Science, Technology, and Cultural Policy and NanoBank, January 17, 2007.) Certain data included herein are derived from the Science Citation Index Expanded, Social Sciences Citation Index, and Arts & Humanities Citation Index of the Institute for Scientific Information®, Inc. (ISI®), Philadelphia, Pennsylvania, USA: © Copyright Institute for Scientific Information®, Inc. 2006. All rights reserved.

## References

Adamic, L. A. (n.a.). *Zipf, power-laws, and Pareto - a ranking tutorial*. Retrieved March 2, 2010 from http://www.hpl.hp.com/research/idl/papers/ranking/ranking.html

Bookstein, A. (1976). The informetric distributions. *Library Quarterly, 46*(4), 416-423.

Bookstein, A. (1990). Informetric distributions, Part I: Unified overview. *Journal of the American Society for Information Science, 41*(5), 368-375.

Bradford, S. C. (1934). Sources of information on specific subjects. *Engineering, 137*, 85-86.

Burrell, Q. L. (1991). The Bradford distribution and the Gini index. *Scientometrics, 21*(2), 181-194.

Burrell, Q. L. (2008). Extending Lotkian informetrics. *Information Processing & Management, 44*(5), 1794-1807.

Clauset, A., Shalizi, C. R., & Newman, M. (2009). Power-law distributions in empirical data. *Society for Industrial and Applied Mathematics Review, 51*, 661-703.

Csányi, G., & Szendroi, B. (2004). Structure of a large social network. *Physical Review E, 69*(3), 036131.

de Bellis, N. (2009). *Bibliometrics and Citation Analysis: From the Science Citation Index to Cybermetrics*. Lanham: Scarecrow Press.

Egghe, L. (1985). Consequences of Lotka's law for the law of Bradford. *Journal of Documentation, 41*(3), 173-189.

Egghe, L. (2005). *Power Laws in the Information Production Process: Lotkaian Informetrics*. Amsterdam: Elsevier Academic Press.

Glänzel, W. (2007). Characteristic scores and scales: A bibliometric analysis of subject characteristics based on long-term citation observation. *Journal of Informetrics, 1*(1), 92-102.

Golder, S. A., & Huberman, B. A. (2006). Usage patterns of collaborative tagging systems. *Journal of Information Science, 32*(2), 198-208.

Leimkuhler, F. F. (1967). The Bradford distribution. *Journal of Documentation, 23*(3), 197-207.

Lotka, A. J. (1926). The frequency distribution of scientific productivity. *Journal of the Washington Academy of Sciences, 16*, 317-323.

Milojević, S. (2010). Modes of collaboration in modern science - Beyond power laws and preferential attachment. *Journal of the American Society for Information Science and Technology*. 61(7) : 1410-1423.

Newman, M. E. J. (2005). Power laws, Pareto distributions and Zipf's law. *Contemporary Physics, 46*(5), 323-351.

Nicolaisen, J., & Hjørland, B. (2007). Practical potentials of Bradford's law: A critical examination of the received view. *Journal of Documentation, 63*(3), 359-377.

Pao, M. L. (1986). An empirical examination of Lotka's law. *Journal of the American Society for Information Science, 37*(1), 26-33.

Price, D. J. d. S. (1965). Networks of scientific papers. *Science, 149(30 July)*, 510-515.

Redner, S. (1998). How popular is your paper? An empirical study of the citation distribution. *European Physical Journal B, 4*(2), 131-134.

Rousseau, B., & Rousseau, R. (2000). LOTKA: A program to fit a power law distribution to observed frequency data. *Cybermetrics, 4*(1), paper 4.

Rousseau, R. (1993). A table for estimating the exponent in Lotka's law. *Journal of Documentation, 49*(4), 409-412.

Zipf, G. K. (1949). *Human Behavior and the Principle of Least Effort*. Cambridge: Addison Weasley.

Zucker, L. G., & Darby, M. R. (2007). Nanobank Data Description release 2.0 (beta-test). Los Angeles, CA: UCLA Center for International Science, Technology and Cultural Policy and Nanobank. January 17, 2007-February 2, 2009.